\documentclass{article}
\usepackage{amsmath}
\usepackage{amsfonts}
\title{Field Identifications for Interacting Bosonic Models in $N=2$ Superconformal Field Theory}
 
\author{Joseph Conlon \footnote{email: jpc41@cam.ac.uk} \\ Christ's College, Cambridge, CB2 3BU, United Kingdom 
\and Doron Gepner \footnote{email: fngepner@wisemail.weizmann.ac.il}
\\Department of Particle Physics, Weizmann Institute of Science,\\ P.O.Box 26, 76100 Rehovot, Israel}

\begin{document}
\maketitle

\begin{abstract}
We study a family of interacting bosonic representations of the $N=2$ superconformal algebra. These models can
be tensored with a conjugate theory to give the free theory. We explain how to use
free fields to study interacting fields and their dimensions, and how we may
identify different free fields as representing the same interacting field. We show how a lattice of identifying fields
may be built up and how every free field may be reduced to a standard form, thus permitting the resolution of the spectrum. 
We explain how to build the extended algebra and show that there are a finite number of primary fields for
this algebra for any of the models.
We illustrate this by studying an example.
\end{abstract}

Conformal field theory has received much attention since the 1984 paper of Belavin, Polyakov
and Zamolodchikov\cite{Belavin}. The addition of $N$ supersymmetry generators extends the
conformal algbera to the level-$N$ superconformal algebra and $N=2$ superconformal field theory is
of interest in the compactification of string theory with $N=1$ space-time supersymmetry. \cite{Trieste} Two representations of 
the $N=2$ algebra were constructed by Kazama and Suzuki \cite{Kazama1}\cite{Kazama2}, and the $G/H$ method has since 
been used extensively. We focus on the representation based on interacting bosons, which contains the $G/H$ models and which
has been studied by Gepner and Cohen \cite{Gepner}\cite{Cohen}. Similar representations can be constructed for $N=0$
conformal field theory\cite{Kiritsis}.
It was shown in \cite{Gepner} that these models can be solved by writing the free theory as a tensor
product of the interacting theory with a conjugate one, allowing free fields to be used to study the dimensions of the interacting theory.
However,
free fields differing only by a field in the conjugate theory correspond to the same interacting field and must
be identified. 
In section \ref{sec:model} we summarise the properties of the models and the relevant results of \cite{Kazama2}\cite{Gepner}\cite{Cohen}. 
In section \ref{Fields} we describe how
the interacting fields and their dimensions can be found through the construction of a mixing matrix.
 In section \ref{Identifying Fields} we address the question of field identifications, and show how a
lattice of identifying fields may be constructed which enables the reduction of every field to a standard form, permitting the solution 
of the spectrum. In the appendix we give the detailed rules for constructing the mixing matrix.

\section{The Models}
\label{sec:model}

The fields are constructed from a vector of bosons $\phi = (\phi_1, \phi_2, \ldots, \phi_n)$ with
standard OPE $\phi_i(z) \phi_j(w) \sim -\delta_{ij} \log(z - w)$. Suppose we have a set of
vectors $\{\gamma_1, \ldots \gamma_n \}$ such that 
\begin{equation}
\Gamma_{ij} = \gamma_i \cdot \gamma_j =  \left\{ \begin{array}{ll} 3 & i=j \\ 0 \textrm{ or } 1 & i \ne j \end{array} \right.
\end{equation}
and a vector $x$ satisfying
\begin{equation}
\label{Gammax=2}
\sum_j \Gamma_{ij}x_{j} = 2
\end{equation}
Then a unitary representation of the $N=2$ superconformal algebra is \cite{Kazama2}
\begin{eqnarray}
\label{qnumber}
J(z) &  = & iq\cdot\partial\phi(z) \\ 
G_{+}(z) & = & \sum_j g_j \exp(i\gamma_j \cdot \phi(z)) \\
G_{-}(z) & = & \sum_j g_j^{\dagger} \exp(-i\gamma_j \cdot \phi(z)) \\
T_A(z) & = & -\frac{1}{4} \sum_i x_i : ( \gamma_i \cdot \partial\phi(z) ) ^2 : + \frac{1}{2} \sum_{\Gamma_{ij}=1} g_i g_j^{\dagger} 
\exp\left(i(\gamma_i - \gamma_j)\cdot\phi(z)\right)
\label{stressT}
\end{eqnarray}

$g_i$ are c-numbers containing a cocycle factor and satisfying $|g_i|^2 = x_i$.
The cocycle factors will be crucial to our subsequent analysis and are further discussed in section \ref{Fields}.
$q$ is given by
\begin{equation}
q = \frac{1}{2} \sum_j x_j \gamma_j
\end{equation}

Note that $q_i \equiv q\cdot\gamma_i = 1$. All properties of the model are determined by $\Gamma$ and $x$
and are independent of the details of $\gamma_i$.  If $\Gamma$ is non-singular, equation (\ref{Gammax=2}) has a unique solution
$x_j = 2 \sum_k \Gamma_{jk}^{-1}$ and the model is called regular. 
If $\Gamma$ is singular equation (\ref{Gammax=2}) has an infinite number of solutions. Singular models
have a flat direction along which the fields may be continuously deformed without
changing the central charge.
We will only consider regular models; however much of our discussion should apply to singular models.

The central charge is given by
\begin{equation}
c = \frac{3}{2}\sum x_i
\end{equation}
It  was found in \cite{Gepner} that $\Gamma$ can be chosen to give rise to 
theories having every possible rational value of the central charge. In
particular, the $k$th minimal model is generated  by the $k$th complete $\Gamma$ matrix 
($\Gamma_{ij} = 1, i \ne j, i,j = 1 \ldots k)$.

The solution of these models depends on an insight. The free stress-energy tensor 
$T_f(z) = -\frac{1}{2} : \partial \phi(z) ^2: $ can be written as
\begin{equation}
T_f(z) = T_A(z) + T_B(z)
\end{equation}
where $T_B = T_f - T_A$ is the conjugate stress tensor. 
As $[T_A, T_B] = 0$, the free theory is a product of two 
distinct theories, $A$ and $B$, and we can write free fields $\Phi_f$ as
\begin{displaymath}
\Phi_f = \Phi_A \otimes \Phi_B
\end{displaymath}
Consequently we can use free fields to study the fields and dimensions of the interacting model.
However, it behoves us to remember that different free fields may correspond to the same
interacting field.
\begin{displaymath}
\Phi_f^{'} = \Phi_A \otimes \Phi_B^{'} \textrm{ and } \Phi_f^{''} = \Phi_A \otimes \Phi_B^{''}
\end{displaymath}
both represent the same interacting field $\Phi_A$. The problem of how to identify $\Phi_f^{'}$ and 
$\Phi_f^{''}$ as the the same interacting field is the field identification problem and will be
studied in section \ref{Identifying Fields}.

\section{The Fields}
\label{Fields}

The dimension $h_A$ of a field $\phi$ is defined by
\begin{equation}
\label{fielddim}
T_A(z)\phi(w) \sim \ldots + \frac{h_A\phi(w)}{(z-w)^2} + \ldots
\end{equation}
where the first dots represent terms of $O(z-w)^{-3}$ and larger, and the second terms of $O(z-w)^{-1}$ and smaller.
Using free fields $\Phi_f(w)$ in (\ref{fielddim}), the formal behaviour for a primary field of the interacting theory is
\begin{eqnarray}
T_A(z)\Phi_f(w) & = & \left(T_A(z)\otimes 1\right)\left(\Phi_A(w) \otimes \Phi_B(w)\right) \nonumber\\
& = & \left(\frac{h_A\Phi_A(w)}{(z-w)^2} + \ldots \right) \otimes \Phi_B(w) = \frac{h_A\Phi_f(w)}{(z-w)^2} + \ldots
\end{eqnarray}

The (not necessarily primary) fields $\Phi_f(w)$ of the free theory can be constructed 
from any combination of $\alpha\cdot\partial \phi(w), \beta\cdot\partial^2 \phi(w), \dots$ and $e^{ip\cdot\phi(w)}$.
However, we are restricted by a locality condition with $G_{\pm}(z)$ which requires that
\begin{equation}
\label{Locality}
\epsilon_i \equiv p \cdot \gamma_i \in \mathbb{Z} 
\end{equation}
This restricts $p = \sum \epsilon_i \zeta_i$ to the dual lattice $\gamma^{*}$ generated by $\zeta_i$, where $\zeta_i \cdot \gamma_j = \delta_{ij}$. 
The $U(1)$ charge $Q$ is easily calculated. 
For a field of the form $\hat{D}e^{ip\phi(w)}$, where $\hat{D}$ are derivative terms as discussed above, e.g.
$\hat{D}^{'} = (\alpha\cdot\partial^4\phi(w))(\beta \cdot \partial \phi(w))$, we have
\begin{equation}
\label{Qvalue}
Q = q \cdot p = \sum \frac{\epsilon_i x_i}{2}
\end{equation}
Note that $Q$ is unaffected by derivative terms.

In the case that $\epsilon_i \in \{0,1\}$ and $\hat{D} = 1$, the field is chiral. However,
free fields do not in general have well-defined dimensions in the interacting theory
as they are mixed by the exponential terms in $T_A(z)$. 
This mixing does not affect $Q$ as $q_i - q_j = 0$. 
By repeated mixing, we can generate a closed set of
free fields $\{ \Phi_1, \Phi_2, \ldots \, \Phi_l\}$, such that
\begin{equation}
T_A(z)\Phi_i(w) = \ldots + \frac{\sum c_{ij} \Phi_j(w)}{(z-w)^2} + \ldots
\end{equation}
The coefficients $c_{ij}$ make up the mixing matrix $M$, whose eigenvectors and eigenvalues give
the fields and dimensions of the interacting model. 
It is found empirically that the dimensions are always rational\cite{Gepner}. As
there are a finite number of fields with a given free dimension that satisfy the locality condition
(\ref{Locality}), the mixing process must terminate
and the mixing matrix is always finite. 

In principle, we can construct a mixing matrix starting from any free field. However, this is not practically feasible.
Consider 
\begin{displaymath}
e^{i(\gamma_i - \gamma_j)\phi(z)} e^{ip\phi(w)} = (z - w)^{\epsilon_i - \epsilon_j} : e^{i(\gamma_i - \gamma_j)\phi(z) + ip\phi(w)} :
\end{displaymath}
For $\epsilon_i - \epsilon_j < -2$ and $\Gamma_{ij}=1$, the $O(z-w)^{-2}$ contribution is calculated by Taylor-expanding the 
$z$-dependent part of the exponential about $w$. This brings down derivatives $i(\gamma_i - \gamma_j)\cdot\partial\phi(w)$,
$i(\gamma_i - \gamma_j)\cdot\partial^2\phi(w)$, $\ldots$ alongside the vertex operator. 
Fields exist containing an arbitrary number of derivatives, but for large negative values of $\epsilon_i - \epsilon_j$,
the number of terms one must consider grows exponentially. In \cite{Gepner} and \cite{Cohen}, only
fields consisting of vertex operators were considered,
which is equivalent to the restriction $\epsilon_i - \epsilon_j \ge -2$
 whenever $\Gamma_{ij} = 1$.
We have extended this treatment to include fields involving up to two derivatives.
For generating vertex operators, the equivalent restriction is now
$\epsilon_i - \epsilon_j \ge -4$. 
This extension is essential for the discovery of identity fields and the subsequent resolution
of the spectrum. 

In our treatment four types of fields may appear:
\newline

\begin{tabular}{lr}
Type 1: & $e^{ip.\phi(w)} $ \\
Type 2: &$ik\cdot\partial\phi(w)e^{iq.\phi(w)}$ \\
Type 3: &$ik\cdot\partial^2\phi(w)e^{ir.\phi(w)}$ \\
Type 4: &$\alpha.\partial\phi(w)\beta.\partial\phi(w)e^{is.\phi(w)}$ 
\end{tabular}
\newline

The free dimensions of these fields are
respectively $D$, $D-1$, $D-2$ and $D-2$.
There are two parts of $T(z)$. $-\frac{1}{4} \sum_i x_i : ( \gamma_i . \partial\phi(z) ) ^2 :$
does not interconvert fields among different types, whereas $\frac{1}{2} \sum g_i g_j^{\dagger} e^{i(\gamma_i - \gamma_j).\phi(z)}$
does. To construct the mixing matrix we must examine the OPEs of each
part of $T(z)$ with all four types of field. We do this explicitly for one case and consign the rest to the appendix. Consider 
\begin{eqnarray}
& e^{i(\gamma_i - \gamma_j)\cdot\phi(z)} (ik\cdot\partial^2\phi(w))e^{ip\cdot\phi(w)} \nonumber\\
= & \left( -\frac{(\gamma_i - \gamma_j)\cdot k}{(z-w)^2} + ik\cdot\partial^2\phi(w) \right)e^{i(\gamma_i - \gamma_j)\cdot\phi(z)}e^{ip\cdot\phi(w)} \nonumber\\
= & \left( \frac{k_j - k_i}{(z-w)^2} + ik\cdot\partial^2\phi(w) \right) (z-w)^{\epsilon_i - \epsilon_j} e^{i((\gamma_i - \gamma_j)\cdot\phi(z) + p\phi(w))} \nonumber
\end{eqnarray}

Taylor-expanding about $z$, we see that
the mixing rules for a type 3 field with the exponential part of $T(z)$ can be written
\newline
\newline
\begin{tabular}{l | c | r}
Condition & Mixes to & Coefficient \\
\hline
$ \epsilon_i - \epsilon_j = 0 $ & $ c_{\gamma_i}c_{-\gamma_j}e^{i(p + \gamma_i - \gamma_j)\cdot\phi(w)} $ & $\frac{1}{2}g_i g_j (k_j - k_i) $\\
\hline
$ \epsilon_i - \epsilon_j = -1 $&$ c_{\gamma_i}c_{-\gamma_j}i(\gamma_i - \gamma_j)\cdot\partial\phi(w)e^{i(p + \gamma_i - \gamma_j).\phi(w)}$&$\frac{1}{2}g_i g_j (k_j - k_i)$ \\
\hline
$\epsilon_i - \epsilon_j = -2 $&$ c_{\gamma_i}c_{-\gamma_j}ik\cdot\partial^2\phi(w)e^{i(p + \gamma_i - \gamma_j)\cdot\phi(w)}$ &$ \frac{1}{2} g_i g_j $\\
&$ c_{\gamma_i}c_{-\gamma_j}i(\gamma_i - \gamma_j)\cdot\partial^2\phi(w)e^{i(p + \gamma_i - \gamma_j)\cdot\phi(w)}$ & $\frac{1}{4}g_i g_j (k_j - k_i) $\\
&$c_{\gamma_i}c_{-\gamma_j}((\gamma_i - \gamma_j)\cdot\partial\phi(w))^2 
e^{i(p+\gamma_i - \gamma_j )\cdot\phi(w)}$ &$-\frac{1}{4} g_i g_j(k_j - k_i)$\\
\end{tabular}
\newline

It is as easy to see in principle as it is tedious in practice how similar expressions can be drawn up for other fields. 
If we
represent the original field by $\Phi_i(w)$, then the new fields produced are in the set $\{ \Phi_1, \Phi_2 \ldots \}$ and
the mixing coefficents correspond to the elements $c_{ij}$ of the mixing matrix. 

Here we must consider
the neglected cocycle factors, $c_{\gamma_i}$ and $c_{\gamma_j}$.  These appear explicitly as part of the fields in the table above.
$g_i$ and $g_j$ have been divested of their cocycle factors and should be treated as c-numbers.
The cocycle factors satisfy \cite{Kazama2}
\begin{eqnarray}
c_{\gamma_1}c_{\gamma_2} & = & \epsilon(\gamma_1 , \gamma_2) c_{\gamma_1 + \gamma_2} \nonumber\\
\epsilon(\gamma_1 , \gamma_2) & = & (-1)^{\gamma_1 \cdot \gamma_2 + \gamma_1^2 \gamma_2^2}\epsilon(\gamma_2 , \gamma_1) \nonumber\\
\epsilon(\gamma_1 , \gamma_2)\epsilon(\gamma_1 + \gamma_2, \gamma_3) & = & \epsilon(\gamma_1 , \gamma_2 + \gamma_3)\epsilon(\gamma_2 , \gamma_3) \\
c_{\gamma}c_{-\gamma} & = & c_{-\gamma}c_{\gamma} = 1 \nonumber\\
c_{\gamma}^{\dagger} & = & c_{-\gamma} \nonumber
\end{eqnarray}
For us, the pertinent relations are the first two. As $\gamma_i^2 = 3$, these imply that
\begin{equation}
c_{\gamma_1} c_{\gamma_2} = (-1)^{1 + \gamma_1 \cdot \gamma_2} c_{\gamma_2} c_{\gamma_1}
\end{equation}
Thus, if $\gamma_1 \cdot \gamma_2 = 0$ the cocycle factors anticommute and hence it is necessary
to explicitly retain them in the construction of the mixing matrix.

We have written a MATLAB program \footnote{available on request from JC} which incorporates the above rules and automatically generates and diagonalises 
the mixing matrix.
Using our program, we can search for fields of the interacting theory containing up to 
two derivatives. Our confidence in the program is assured by the fact that all the dimensions produced were rational,
a fact which breaks down immediately in the presence of errors.

\section{Field Identifications}
\label{Identifying Fields}

The methods of section \ref{Fields} enable us to generate a plethora of fields having 
well-defined interacting dimensions and $U(1)$ charges. 
To solve the interacting theory, it is necessary to identify the spectrum
of fields and dimensions. The major problem is knowing when different free fields $\Phi_f$ and $\Phi_f^{'}$ correspond to the 
same interacting field.
In the case of minimal models, we know what the field spectrum must be and we can cheat. 
In the general case, a more principled approach is required.
Identifications can be made through fields entirely in the conjugate theory, which correspond to the identity in the 
$A$ theory. These are characterised by having $Q=0$ and $h_A=0$. If $\Phi_I(z)$ is one such field, then we have
\begin{eqnarray}
\Phi_I(z) \Phi_f(w) & = & (1 \otimes \Phi^{'}_B(z))(\Phi_A(w)\otimes\Phi_B(w)) \nonumber\\
& = & (z-w)^{-\alpha}(\Phi_A(w)\otimes \Phi_B^{''}(w) + \ldots ) \nonumber
\end{eqnarray}
This allows the identification of $\Phi_A \otimes \Phi_B$ with $\Phi_A \otimes \Phi_B^{''}$.
The discovery of all identity fields
is necessary for the solution of the spectrum.
A field identification existing for all models is $T_B(z)$, as it lies purely in the conjugate theory. Another easy source
of identity fields are those due to symmetries. If $\Gamma_{ij}=1$ and the interchange of $\gamma_i$ and $\gamma_j$ leaves $\Gamma$ unchanged,
then $e^{i(\zeta_i - \zeta_j)\cdot\phi(w)} - c_{\gamma_i}c_{-\gamma_j}e^{i(\zeta_j - \zeta_i)\cdot\phi(w)}$ will always be an identifying field. These
do not exhaust the supply of identifying fields - some can only be found by automated search,
as we shall see by studying an example. It should be noted that in general field identifications, e.g. $T_B(z)$, require derivative terms.

Two identity fields can reproduce through their OPE.
\begin{displaymath}
I_1(z)I_2(w) = (1\otimes \Phi_B(z))(1 \otimes \Phi_B^{'}(w)) = (z-w)^{\alpha}(1 \otimes \Phi_B^{''}(w) + \ldots )
\end{displaymath}
Repeated use of such OPEs allows the generation of an infinite lattice, $I^{*}$, of identity fields, the lattice points
corresponding to the exponents that are present. 
\begin{displaymath}
\underbrace{(e^{ip_I\phi(z)} + \ldots)}_{I_p}\underbrace{(e^{iq_I\phi(w)} + \ldots )}_{I_q} = 
(z-w)^{p_I \cdot q_I}\underbrace{(e^{i(p_I + q_I)\phi(w)} + \dots )}_{I_{p+q}}
\end{displaymath}
We note that it is not in general the case
that the dimension of the lattice $I^{*}$ equals the number of independent identity fields. For example, $T_B(z)$ will
often contain several independent exponents. 

The lattice $I^{*}$ is the key to reducing the number of fields to a manageable amount. As $\gamma^* / I^*$ is discrete, it
is unchanged by continuous deformations of the theory - i.e. it is topological. Treating $\gamma^* / I^*$ as a set of 
`standard exponents', there are two steps in the simplification process.
First, an arbitrary field $F_p(w) = (\hat{D}e^{ip\phi(w)} + \ldots)$
is identified with a field containing the standard exponent $p+s$.
\begin{displaymath}
\underbrace{(e^{is_I\phi(z)} + \ldots )}_{I_s}\underbrace{(\hat{D}(l)e^{ip\phi(w)} + \ldots)}_{F_p} 
= \frac{(z-w)^{s_I \cdot p}}{(z-w)^{l-m}} \big(\underbrace{\hat{D}(m) e^{i(p+s)\phi(w)} + \ldots}_{F_{p+s}} \big)
\end{displaymath}
Secondly, an identifying field containing a zero-momentum term is used to reduce the number of
derivatives as much as possible. Examples of such fields are $T_B(z)$ and those generated from the OPE $I(z)I^{\dagger}(w)$,
where $I$ is an identifying field.
\begin{equation}
\label{wrongequation}
\underbrace{(\hat{D}(k) + \ldots )}_{I_0}\underbrace{(\hat{D}(m)e^{i(p+s)\phi(w)} + \ldots)}_{F_{p+s}} 
= \frac{1}{(z-w)^{\alpha}} \big(\underbrace{e^{i(p+s)\phi(w)} + \ldots}_{F'_{p+s}} \big) 
\end{equation}
Equation (\ref{wrongequation}) is not necessarily correct, as in general it is not possible to remove all derivatives from the standard exponent $p+s$.
We consider the set $\{ \xi \in \mathbb{R}^n | \xi \cdot I^* = 0\}$, with generators $S = \{q_1 \ldots q_n\}$. $S$ always
contains the $U(1)$ current vector $q$ of equation ($\ref{qnumber}$) and, in the case where $(\gamma^* / I^*)|_{Q=0}$ is of size one, this is the only element.
The identifying fields with zero momentum are generated by expressions of the type $\prod_{i,j} (p_i \cdot \partial^j \phi)$, where
$p_i \in I^*$. This follows from the OPE $I_n(z)I_n^{\dagger}(w) = \prod_{i, j} A_{ij} 
(p_{i}\cdot\partial^{j}\phi) + \ldots$, where $p_{\alpha} \in I^*$. 
If we have a term $\prod_{\alpha, \beta} (q_\alpha \cdot \partial^{\beta} \phi(w))$, where $q_{\alpha} \in S$, then its OPE
with an identifying field is non-singular by Wick's theorem, as we have
\begin{displaymath}
p_i \cdot \partial^j \phi(z) q_\alpha \cdot \partial^{\beta} \phi(w) \propto p_i \cdot q_{\alpha}  = 0
\end{displaymath}
Any derivative term can be written 
\begin{displaymath}
\hat{D} = \underbrace{\prod_{i,j} (p_i \cdot \partial^j \phi)}_{\hat{D}_1} \underbrace{\prod_{\alpha, \beta} (q_\alpha \cdot \partial^{\beta} \phi)}_{\hat{D}_2}
\end{displaymath}
The field $\hat{D}$ does not depend on $D_1$ as it is a field identification and we can assume $\hat{D}_2 = 1$.
Consequently, $\hat{D}_1$-style derivative terms can be eliminated and we are left with those in the form of $\hat{D}_2$.
Hence equation (\ref{wrongequation}) should instead read
\begin{eqnarray}
\label{rightequation}
\left( \hat{D}_1^{'}(k) + \ldots \right) & \left(\prod_{i,j} (p_i \cdot \partial^j \phi) \prod_{\alpha,\beta} 
(q_{\alpha} \cdot \partial^{\beta} \phi)(e^{i(p+q)\phi(w)} + \ldots )\right) = \nonumber\\
& \frac{1}{(z-w)^{\gamma}} \left( \prod_{\alpha,\beta} (q_{\alpha} \cdot \partial^{\beta} \phi) e^{i(p+q)\phi(w)} + \ldots \right) 
\end{eqnarray}
   
The procedure for the solution of the models is now clear. Having searched for and found the field identifications, we construct the set of standard exponents
$\gamma^{*}/I^{*}$. Any field $\hat{P}$ generated by the mixing matrix is associated with a standard exponent $p$ and 
can be identified with a field containing terms in $p$. We can use field identifications as in equation (\ref{rightequation})
to define a standard form for each field,
which is the identified field which contains $p$ and whose derivative terms are solely those of the form 
$\prod_{\alpha,\beta} (q_\alpha \cdot \partial^{\beta} \phi(w))$ where $q_{\alpha} \in S$. To find the spectrum 
of fields, we start with each standard exponent and examine the fields generated by the mixing process.
We then add derivative terms $q_i \cdot \partial\phi(w)$, mix the new fields, and so on. The spectrum consists of all
independent fields thus produced.

The set  $S$ always contains the $U(1)$ current vector
$q$ of equation (\ref{qnumber}). It can also contain other vectors $q_{\alpha}$. These vectors can be used
to enlarge the algebra - i.e. adding $q_{\alpha} \cdot \partial \phi(w)$ as generators. This means that the
derivative fields will always be descendants of this larger algebra. In addition, we can also add the fields
$e^{ip\phi(w)}$, where $p \in P$, where $P = \{ x | x \in \gamma^*,  x \cdot I^* = 0\}$. These fields will be local w.r.t. $G_{\pm}$ as they belong to 
$\gamma^*$. It can be seen that with these extensions we will have a finite number of primary fields modulo the extended algebra, and so 
this is a rational conformal field theory with this extended algebra. The set of primary momenta $\gamma^* / (P \oplus I^*)$ is finite because
dim($P\oplus I^*$) = dim($\gamma^*$). it can be seen that these generators of the extended algebra lie entirely in the $A$ theory i.e. 
their naive dimensions are equal to their actual dimensions. This guarantees that this is the correct algebra. 

One question we have not been able to solve is how to tell when all independent identity fields have been found. This can be answered
affirmatively if the lattice $I^*$ is the same as $\gamma^*|_{Q=0}$, and may often be answered if the model is a tensor product of distinct theories. 
An example is the $k=28$ minimal model generated by 
\begin{displaymath}
\Gamma = \left( \begin{array}{cccc} 3 & 1 & 1 & 0 \\ 1 & 3 & 1 & 0 \\ 1 & 1 & 3 & 0 \\ 0 & 0 & 0 & 3 \end{array} \right)
\end{displaymath}
Here the only field identifications needed are those for the complete $3\times3$ $\Gamma$ matrix. We also note
that in this case $\textrm{dim}(I^*) = 2 < 3 = \textrm{dim}(\gamma^* | _{Q=0})$. In view of the above remark on
extended algebras, this is consistent as this theory can be written $(N=2) \oplus (N=2)$. However, for general non-minimal models this 
question is not easy and is related
to that of how to know in advance the form of the identifying fields. In all minimal models that we have studied,
two derivatives are sufficient to find a complete set of generators for the lattice $I^{*}$. As the number of derivatives involved 
is a discontinuous quantity, we expect the structure of the identifying fields to be unchanged for models related by a process 
of continuous deformation.

As an illustration we resolve an example of Cohen.
Consider the model generated by 
\begin{displaymath}
\Gamma = \left( \begin{array}{cccc} 3 & 1 & 1 & 0 \\ 1 & 3 & 1 & 0 \\ 1 & 1 & 3 & 1 \\ 0 & 0 & 1 & 3 \end{array} \right)
\end{displaymath}
This model has $\frac{c}{3} = \frac{11}{13}$ and corresponds to the $k=11$ minimal model. We have
\begin{displaymath}
\Gamma^{-1} = \frac{1}{52} \left( \begin{array}{cccc} 21 & -5 & -6 & 2 \\ -5 & 21 & -6 & 2 \\ -6 & -6 & 24 & -8 \\ 2 & 2 & -8 & 20 \end{array} \right)
\end{displaymath}
$x$ is given by
$(x_1, x_2 , x_3, x_4) = \frac{1}{13}(6, 6, 2, 8)$ and the $U(1)$ charges are in units of $\frac{1}{13}$. Particularly interesting
are the chiral fields, for which $Q=2h$ and which have charges $Q = 0 \to \frac{11}{13}$. These are
\begin{eqnarray}
\Upsilon_0 & = & 1 \nonumber\\
\Upsilon_1 & = & e^{\zeta_3} \nonumber\\
\Upsilon_2 & = & e^{2\zeta_3} + e^{2\zeta_2 - \zeta_4} + e^{2\zeta_1 - \zeta_4} + e^{-\zeta_2 + 2\zeta_4 - \zeta_1} \nonumber \\
\Upsilon_3 & = & e^{\zeta_1} \nonumber\\
\Upsilon_4 & = & e^{\zeta_4} \nonumber\\
\Upsilon_5 & = & e^{\zeta_3 + \zeta_4} \nonumber\\
\Upsilon_6 & = & e^{\zeta_1 + \zeta_2} \nonumber\\
\Upsilon_7 & = & e^{\zeta_1 + \zeta_4} \nonumber\\
\Upsilon_8 & = & e^{\zeta_1 + \zeta_3 + \zeta_4} \nonumber\\
\Upsilon_9 & = & e^{\zeta_1 + \zeta_2 - \zeta_3 + \zeta_4} + e^{\zeta_1 - \zeta_2 + \zeta_3 + 2\zeta_4} + e^{-\zeta_1 + \zeta_2 + \zeta_3 + 2\zeta_4}
             + e^{2\zeta_1 + 2\zeta_2 + \zeta_3 - \zeta_4} \nonumber\\
\Upsilon_{10} & = & e^{\zeta_1 + \zeta_2 + \zeta_4} \nonumber\\
\Upsilon_{11} & = & e^{\zeta_1 + \zeta_2 + \zeta_3 + \zeta_4}\nonumber
\end{eqnarray} 

The problem is that these are not the only chiral fields that can be found.
For example, there are five $Q=\frac{6}{13}$ chiral fields consisting solely of vertex operators. 
\begin{eqnarray}
\Phi_1 & = & e^{\zeta_1 + \zeta_2} \nonumber \\
\Phi_2 & = & e^{\zeta_2 - \zeta_3 + \zeta_4} + e^{-\zeta_2 + \zeta_3 + 2\zeta_4} + e^{\zeta_1 + 2\zeta_2 + \zeta_3 - \zeta_4} \nonumber \\
\Phi_3 & = & e^{\zeta_1 - \zeta_3 + \zeta_4} + e^{-\zeta_1 + \zeta_3 + 2\zeta_4} + e^{2\zeta_1 + \zeta_2 + \zeta_3 - \zeta_4} \nonumber \\
\Phi_4 & = & e^{2\zeta_1} + e^{2\zeta_2} \nonumber \\
\Phi_5 & = & e^{2\zeta_1} + e^{2\zeta_2} + e^{2\zeta_3 + \zeta_4} \nonumber
\end{eqnarray}
We have omitted the coefficients of the fields and the $i\phi(w)$ part of the exponential.
As this is a minimal model, we must be able to explicitly identify these fields.
We can immediately write down two identifying fields, $I_1$, from the symmetries of $\Gamma$, and $I_2 \equiv T_B(z)$.
These are
\begin{eqnarray}
I_1 & = & e^{\zeta_1 - \zeta_2} - c_{\gamma_1}c_{-\gamma_2}e^{\zeta_2 - \zeta_1} \\
I_2 & = & -\frac{1}{2}\partial\phi(w)^2 -\frac{1}{4} \sum_i x_i ( \gamma_i \cdot \partial\phi(z) ) ^2 + \frac{1}{2} 
\sum g_i g_j^{\dagger} e^{i(\gamma_i - \gamma_j)\cdot\phi(z)}
\end{eqnarray}
$I_1$ identifies $\Phi_1$ and $\Phi_4$, and $\Phi_2$ and $\Phi_3$. 
$I_2$ identifies $\Phi_4$ and $\Phi_5$. This leaves two unidentified groups, \{$\Phi_1$, $\Phi_4$, $\Phi_5$\} and \{$\Phi_2$, $\Phi_3$\}.
The missing identity field $I_3$ is not obvious. As it must identify $\Phi_1$ and $\Phi_2$, we consider $\Phi_1 \Phi_2^{\dagger}$ and use our program 
to search for an identifying field. It turns out that 
$I_3$ is the sum of the following twelve fields, where [a, b, c, d] denotes $a\zeta_1 + b\zeta_2 + c\zeta_3 + d\zeta_4$. 
\begin{eqnarray}
\Psi_1 & = & e^{[1,0,-3,0]} \nonumber \\
\Psi_2 & = & c_{-\gamma_1}c_{-\gamma_2}c_{\gamma_3}^3 c_{-\gamma_4} \times -e^{[0,-1,3,0]} \nonumber\\ 
\Psi_3 & = & c_{-\gamma_1}^2 c_{\gamma_3}^2 \times \frac{1}{15}e^{[-3,0,1,2]} \nonumber \\
\Psi_4 & = & c_{-\gamma_1}c_{\gamma_2} c_{\gamma_3} c_{-\gamma_4} \times -\frac{1}{15}e^{[0,3,-1,-2]} \nonumber \\
\Psi_5 & = & c_{\gamma_3}c_{-\gamma_4}\times \frac{1}{15}[-2,0,-6,3]\cdot\partial\phi(w) e^{[2,1,-1,-2]} \nonumber\\
\Psi_6 & = & c_{-\gamma_1}c_{-\gamma_2}c_{\gamma_3}^2 \times \frac{1}{15}[0,-2,-6,3]\cdot\partial\phi(w) e^{[-1, -2,1,2]} \nonumber\\
\Psi_7 & = & c_{-\gamma_2}c_{\gamma_3}^2 c_{-\gamma_4}\times \frac{\sqrt{3}}{45}
             [4,2,-6,-3]\cdot\partial\phi(w) e^{[2,-1,1,-1]} \nonumber\\
\Psi_8 & = & c_{-\gamma_2}c_{\gamma_3}\times \frac{\sqrt{3}}{45}[2,4,-6,-3]\cdot\partial\phi(w) e^{[1,-2,-1,1]} \nonumber\\
\Psi_9 & = & c_{-\gamma_1}c_{\gamma_3}\times\frac{\sqrt{3}}{90}[4, 2, -30, 3]\cdot\partial^2\phi(w) e^{[-1,0,-1,1]} \nonumber\\
\Psi_{10} & = & c_{-\gamma_1}c_{\gamma_3}^2 c_{-\gamma_4}\times \frac{\sqrt{3}}{90}[2,4,-30,3]\cdot\partial^2\phi(w) e^{[0,1,1,-1]} \nonumber\\
\Psi_{11} & = & c_{-\gamma_1}c_{\gamma_3}\times\frac{\sqrt{3}}{90}\big( [0,2,-2,-1]\cdot\partial\phi(w)[-4,-6,10,5]\cdot\partial\phi(w) \nonumber\\
          & & + [-2, -2, -4, 4]\cdot\partial\phi(w)[-3,-3,10,2]\cdot\partial\phi(w) \big) e^{[-1,0,-1,1]} \nonumber\\
\Psi_{12} & = & c_{-\gamma_1}c_{\gamma_3}^2 c_{-\gamma_4}\times\frac{\sqrt{3}}{90}\big( [-2,0, 2,1]\cdot\partial\phi(w)[-6,-4,10,5]\cdot\partial\phi(w) \nonumber\\
          & &  + [-2, -2, -4, 4]\cdot\partial\phi(w)[3, 3,-10,-2]\cdot\partial\phi(w) \big) e^{[0,1,1,-1]} \nonumber
\end{eqnarray}

This completes the set of field identifications. In a similar fashion, all chiral fields of given charge can be explicitly identified, resulting in the 
correct spectrum for the $k=11$ minimal model.
The lattice $I^{*}$ produced by $I_1, I_2$ and $I_3$ equals $\gamma^{*}|_{Q=0}$. This can be seen by noting that any
exponent [$a$, $b$, $c$, $d$] which satisfies $Q=0$ must have $c= -3a -3b - 4d$. It is then possible to write
\begin{displaymath}
[a,b,c,d] = b[-1,1,0,0] + (a+b+d)[2,0,-2,-1] + (a+b+2d)[-1,0,-1,1]
\end{displaymath}
As the exponents [-1, 1, 0, 0], [2, 0, -2, -1] and [-1, 0, -1, 1] are part of $I_1$, $I_2$, and $I_3$ respectively, 
an identity field containing the exponent $[a, b, c, d]$  may be constructed through studying terms in the OPE of 
$I_1^b I_2^{a+b+d} I_3^{a+b+2d}$. These identifying fields can be used to reduce any field to a standard form as discussed above. For example, all fields
of charge $Q=\frac{3}{13}$ can be identified with one containing the exponent $\zeta_1$.

We conclude that by using field identifications we are able to correctly identify the spectrum of the interacting models and `factor out' 
the conjugate theory. As well as the above example, we have successfully applied this technique to other minimal models - e.g. 
the $k=35$ minimal model generated by a $5\times5$ $\Gamma$ matrix. In all cases studied, two derivatives have been sufficient to find all independent field
identifications. For $c\ge3$ models, further work is necessary to tackle the problem of how to know when we have the complete set of identifying fields.

\subsection*{Acknowledgments}

JC thanks the Weizmann Institute of Science for a summer studentship and Christ's College, Cambridge for a travel grant.

\appendix
\section*{Appendix}

Here we summarise the rules for generating the mixing matrix. 
\newline

\underline{Mixing due to exponential part of $T(z)$:}
\newline
\begin{tabular}{|l | c | r|}
\hline
Type 1 & $e^{p}$& \\
\hline
Condition & Mixes to ($\times c_{\gamma_i}c_{-\gamma_j}$) & Mixing Coefficient \\
\hline
$ \epsilon_i - \epsilon_j = -2 $ & $ e^{p + \gamma_i - \gamma_j} $ & $\frac{1}{2}g_i g_j$\\
\hline
$ \epsilon_i - \epsilon_j = -3 $&$ i(\gamma_i - \gamma_j).\partial\phi(w)e^{p + \gamma_i - \gamma_j}$&$\frac{1}{2}g_i g_j$ \\
\hline
$\epsilon_i - \epsilon_j = -4 $&$ i(\gamma_i - \gamma_j).\partial^2\phi(w)e^{p + \gamma_i - \gamma_j}$ &$ \frac{1}{4} g_i g_j $\\
&$ c_{\gamma_i}c_{-\gamma_j}((\gamma_i - \gamma_j).\partial\phi(w))^2 e^{p + \gamma_i - \gamma_j}$ & $-\frac{1}{4}g_i g_j$\\
\hline
Type 2 & $ik\cdot\partial\phi(w) e^{p}$& \\
\hline
Condition & Mixes to ($\times c_{\gamma_i}c_{-\gamma_j}$)& Mixing Coefficient \\
\hline
$ \epsilon_i - \epsilon_j = -1 $ & $ e^{p + \gamma_i - \gamma_j} $ & $\frac{1}{2}g_i g_j (k_j - k_i) $\\
\hline
$ \epsilon_i - \epsilon_j = -2 $&$ i(\gamma_i - \gamma_j).\partial\phi(w)e^{p + \gamma_i - \gamma_j}$&$\frac{1}{2}g_i g_j(k_j - k_i)$ \\
&$ ik.\partial\phi(w)e^{p + \gamma_i - \gamma_j}$&$\frac{1}{2}g_i g_j$ \\
\hline
$\epsilon_i - \epsilon_j = -3 $&$ i(\gamma_i - \gamma_j).\partial^2\phi(w)e^{p + \gamma_i - \gamma_j}$ &$ \frac{1}{4} g_i g_j(k_j - k_i) $\\
&$ k.\partial\phi(w)(\gamma_i - \gamma_j).\partial\phi(w) e^{p + \gamma_i - \gamma_j}$ & $-\frac{1}{2}g_i g_j$\\
&$ ((\gamma_i - \gamma_j).\partial\phi(w))^2 e^{p + \gamma_i - \gamma_j}$ & $\frac{1}{4}g_i g_j(k_i - k_j)$\\
\hline
Type 3 & $ik.\partial^2\cdot\phi(w) e^{p}$& \\
\hline
Condition & Mixes to ($\times c_{\gamma_i}c_{-\gamma_j}$)& Mixing Coefficient \\
\hline
$ \epsilon_i - \epsilon_j = 0 $ & $ e^{p + \gamma_i - \gamma_j} $ & $\frac{1}{2}g_i g_j (k_j - k_i) $\\
\hline
$ \epsilon_i - \epsilon_j = -1 $&$ i(\gamma_i - \gamma_j).\partial\phi(w)e^{p + \gamma_i - \gamma_j}$&$\frac{1}{2}g_i g_j(k_j - k_i)$ \\
\hline
$\epsilon_i - \epsilon_j = -2 $&$ i(\gamma_i - \gamma_j).\partial^2\phi(w)e^{p + \gamma_i - \gamma_j}$ &$ \frac{1}{4} g_i g_j(k_j - k_i) $\\
&$ ik.\partial^2\phi(w) e^{p + \gamma_i - \gamma_j}$ & $\frac{1}{2}g_i g_j$\\
&$ ((\gamma_i - \gamma_j).\partial\phi(w))^2 e^{p + \gamma_i - \gamma_j}$ & $\frac{1}{4}g_i g_j(k_i - k_j)$\\
\hline
\end{tabular}
\newline
\begin{tabular}{|l |c| r|}
\hline
Type 4 & $\alpha\cdot\partial\phi(w) \beta\cdot\partial\phi(w) e^{p}$& \\
\hline
Condition & Mixes to ($\times c_{\gamma_i}c_{-\gamma_j}$)& Mixing Coefficient \\
\hline
$ \epsilon_i - \epsilon_j = 0 $ & $ e^{p + \gamma_i - \gamma_j} $ & $-\frac{1}{2}g_i g_j (\alpha_i - \alpha_j)(\beta_i - \beta_j) $\\
\hline
$ \epsilon_i - \epsilon_j = -1 $&$ i(\gamma_i - \gamma_j).\partial\phi(w)e^{p + \gamma_i - \gamma_j}$&$-\frac{1}{2}g_i g_j(\alpha_i - \alpha_j)(\beta_i - \beta_j)$ \\
&$ i\alpha.\partial\phi(w)e^{p + \gamma_i - \gamma_j}$&$\frac{1}{2}g_i g_j(\beta_i - \beta_j)$ \\
&$ i\beta.\partial\phi(w)e^{p + \gamma_i - \gamma_j}$&$\frac{1}{2}g_i g_j(\alpha_i - \alpha_j)$\\
\hline
$\epsilon_i - \epsilon_j = -2 $&$ i(\gamma_i - \gamma_j).\partial^2\phi(w)e^{p + \gamma_i - \gamma_j}$ &$ -\frac{1}{4} g_i g_j(\alpha_i-\alpha_j)(\beta_i-\beta_j) $\\
&$ \alpha.\partial\phi(w)\beta.\partial\phi(w) e^{p + \gamma_i - \gamma_j}$ & $\frac{1}{2}g_i g_j$\\
&$ \alpha.\partial\phi(w)(\gamma_i - \gamma_j).\partial\phi(w) e^{p + \gamma_i - \gamma_j}$ & $-\frac{1}{2}g_i g_j(\beta_i - \beta_j)$\\
&$ \beta.\partial\phi(w)(\gamma_i - \gamma_j).\partial\phi(w) e^{p + \gamma_i - \gamma_j}$ & $-\frac{1}{2}g_i g_j(\alpha_i - \alpha_j)$\\
&$ ((\gamma_i - \gamma_j).\partial\phi(w))^2 e^{p + \gamma_i - \gamma_j}$ & $\frac{1}{4}g_i g_j(\alpha_i - \alpha_j)(\beta_i - \beta_j)$\\
\hline
\end{tabular}
\newline

\underline{Mixing due to derivative terms in $T_A(z)$}
\newline
\begin{tabular}{|l|c|c|}
\hline
Field Type& Mixes To & Mixing Coefficient \\
\hline
$e^p$ & $ e^{p} $ & $\sum \frac{x_i \epsilon_i^2}{4}$ \\
\hline
$ik.\partial\phi(w) e^p$ & $ i\sum_i k_i x_i \gamma_i.\partial\phi(w) e^{p} $ & $\frac{1}{2}$\\
& $ ik.\partial\phi(w) e^{p} $ & $\sum \frac{x_i \epsilon_i^2}{4}$ \\
\hline
$ik.\partial^2\phi(w) e^p$ & $ i\sum_i k_i x_i \gamma_i.\partial^2\phi(w) e^{p} $ & 1\\
& $ ik.\partial^2\phi(w) e^{p} $ & $\sum \frac{x_i \epsilon_i^2}{4}$ \\
\hline
$\alpha.\partial\phi(w)\beta.\partial\phi(w) e^p$ &$ i\sum_i \alpha_i x_i \gamma_i.\partial\phi(w) \beta.\partial\phi(w) e^{p} $ & $\frac{1}{2}$\\
&$ i\sum_i \beta_i x_i \gamma_i.\partial\phi(w) \alpha.\partial\phi(w) e^{p} $ & $\frac{1}{2}$\\
&$ \alpha.\partial\phi(w) \beta.\partial\phi(w) e^{p} $ & $\sum \frac{x_i \epsilon_i^2}{4}$ \\
\hline
\end{tabular}
\newline

\end{document}